\documentstyle[twocolumn,aps,epsf,floats]{revtex}
\begin{document}
\draft
\title{Dynamical Franz-Keldysh effect}

\author
{A. P. Jauho and K. Johnsen}
\address{
Mikroelektronik Centret, Technical University of Denmark, Bldg 345east\\
DK-2800 Lyngby, Denmark
\medskip\\
\date{Submitted to Phys. Rev. Lett., December 5 1995}
\parbox{14cm}{\rm
We introduce and analyze the properties of dynamical Franz-Keldysh
effect, i.e. the change of density-of-states, or absorption spectra,
of semiconductors under the influence of {\it time-dependent} electric
fields.  In the case of a harmonic time-dependence, we predict the
occurence of significant fine structure, both below and
above the zero-field band-gap, which should be experimentally
observable.\smallskip\\
PACS numbers: 71.10.-w,78.20.Bh,78.20.Jq,78.47.+p
\smallskip\\}}
\maketitle
\narrowtext
Almost forty years ago Franz\cite{Franz} and Keldysh\cite{Keldysh1}
pointed out that static electric fields modify the linear optical
properties of bulk semiconductors near the optical absorption edge:
the absorption coefficient becomes finite (though small) for photon
energies below the band gap, and that the absorption coefficient
shows oscillations for energies above the band gap.  Studies of this
kind regained importance with the development of quantum confined
semiconductor structures, and a large body of both experimental
and theoretical knowledge has been developed since mid 80's
\cite{Schmitt-Rink}.  The purpose of this Letter is to point out
that the absorption curve develops additional structure when
the sample is placed in a strong ac-field, such as generated by
a free-electron laser.  As we shall demonstrate below, the
zero-field density of states splits into replicas, with
field-dependent shifts and time-dependent amplitudes.
In contrary to the static case, where it is sufficient to solve
a simple Schr{\"o}dinger equation, the time-dependent case
necessitates the use of a more advanced formalism; in the present
case we apply nonequilibrium Green function techniques.  In order
to present our predictions in a clear-cut manner, we have stripped
the model calculation from complications that may be present
in real samples, such as scattering due to sample nonidealities
and phonons,
or excitonic effects.  Thus, the main motivation of the present
{\it analytic} study is to point out the existence of an effect,
and not address the quantitative details, which can be treated
properly only
with a full-scale numerical calculation based on, for example,
solution of the semiconductor Bloch equations\cite{Kochgroup}.

The calculation of the electroabsorption coefficient $\alpha(\omega)$
for free
carriers is discussed in text-books\cite{Haug-Koch}, and the
result can be expressed as
\begin{equation}\label{alpha}
\alpha(\omega) = \alpha_0 \sum_n |\psi_n({\bf r}=0)|^2
\delta(E_n - \hbar\omega)\;,
\end{equation}
where $\psi_n$ and $E_n$ are the eigenfunctions and eigenvalues
of the electron-hole pair Schr{\"o}dinger equation, respectively.
The important feature of Eq.(\ref{alpha}) for our purposes
is that it states that the absorption coefficient is essentially
given by the joint density-of-states, $\alpha(\omega)\simeq
\rho(\omega) \equiv \sum_n \delta(\omega-E_n)$.  In equilibrium many-body
theory one can express the density-of-states in terms of the
imaginary part of the retarded
Green function, or, equivalently, by the spectral function.  The
important property of the nonequilibrium
theory of Kadanoff-Baym\cite{Kadanoff-Baym} and Keldysh\cite{Keldysh2}
is that in nonequilibrium the diagrammatic structure of the perturbation
theory is the same as in equilibrium, with the proviso that real-time
integrals must be generalized to a complex contour\cite{reviews}.  The
upshot is that the nonequilibrium retarded Green function can be calculated
from
a Dyson equation which is formally identical to the equilibrium case.
Thus, we can analyze the electroabsorption properties of a semiconductor
in an applied time-dependent electric field by evaluating
\begin{eqnarray}\label{rho}
\rho(\omega,T) & = &-{1\over\pi}\sum_{\bf k}{\rm Im}
{\tilde G}^r(\bf k,\omega,T)\nonumber\\
& = & {1\over 2\pi}\sum_{\bf k} {\tilde A}(\bf k,\omega,T)\;.
\end{eqnarray}
Eq.(\ref{rho}) introduces a
{\it generalized density-of-states} function, which is time-dependent
due to the external electric field.  Whether a measurement
probes $\rho(\omega,T)$ or perhaps its suitably weighted time-average
depends on the various time-scales of the particular
experimental set-up;
here we focus on the full time-dependence, because it contains
most information. In (\ref{rho}) we also introduce the gauge-invariant form
${\tilde G}^r(\bf k,\omega,T)$ of the
two-time Green function $G^r({\bf x},{\bf x}',t,t')$, which
we calculate via its Dyson equation.  Effects due
to interactions can be taken into account by introducing a suitable
self-energy.  Under certain conditions the Dyson equation for the
retarded Green function may involve particle densities; in this case
one must also solve the appropriate quantum kinetic equation
for the distribution function.  However, this problem is not
addressed in the present paper.

In an actual calculation the longitudinal electric
field must be introduced by using some explicit gauge; however predictions for
observables must clearly be gauge-independent, and hence the
gauge-invariant formulation is essential.  In general,
the transformation to gauge invariant
\begin{figure}
\epsfxsize=7.5cm
\epsfbox{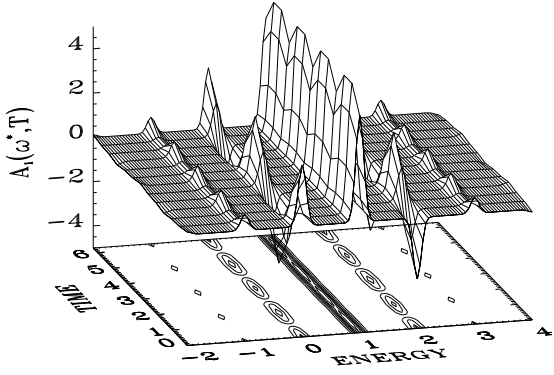}
\vspace{1cm}
\caption{The function ${\tilde A}_1$
as a function of
the variable
$\omega^* = (\omega - \epsilon_k)/\Omega$.  The field strength is given
by $\omega_F/\Omega = 1$.  The $\delta$-function singularities
of ${\tilde A}_1$ have been broadened for illustrative purposes}
\label{f1}
\end{figure}
\noindent
variables occurs via\cite{Rita}
\begin{eqnarray}\label{gaugeinv}
{\tilde G}^r(p,X) & = & \int  d^4 q G^r(q,X)\nonumber\\
&\times& \exp\bigl( i q_{\mu}[p_{\mu} +
\int_{-1/2}^{1/2} d\lambda A_{\mu}(X+\lambda q)]\bigr)
\;,
\end{eqnarray}
where we have used the Einstein
summation convention and a ``covariant" notation, where four-momenta
and coordinates
are given by
$p\equiv(\omega,{\bf k}), q\equiv(\tau,{\bf r})=(t-t',{\bf x}-{\bf x}')$,
and $X\equiv(T,{\bf R})=([t+t'])/2,[{\bf x}+{\bf x}']/2)$.
Correspondingly, the potentials defining the electric field
are contained in
$A\equiv(\phi,{\bf A})$.

In our calculation
of the density-of-states for a uniform
but time-dependent field it is convenient
to describe the electric field with a vector
potential,
${\bf A}(t) = -{\bf E}\sin(\Omega t)/\Omega$ (here
we restrict ourselves to harmonic time-dependence).
The retarded Green function satisfies\cite{dimensions}
\begin{equation}\label{Dyseq}
\Bigl\{i{\partial\over\partial t} -
\epsilon\bigl[{\bf p} - {\bf A}(t)\bigr]\Bigr\}G^r({\bf p},t,t') =
\delta(t-t')\;.
\end{equation}
Using (\ref{gaugeinv}) Eq.(\ref{Dyseq}) immediately leads to the
gauge invariant spectral function ${\tilde A}=i({\tilde G}^r
-{\tilde G}^a)$,
\begin{eqnarray}\label{A}
{\tilde A}({\bf k},\omega,T) &=&
\int d{\bf r}d \tau e^{i w}
\int d{\bf p} e^{i{\bf r}\cdot{\bf p}}\nonumber\\
&\times& \exp\Bigl\{-i
\int_{T-\tau/2}^{T+\tau/2} d t_1 \epsilon\bigl[{\bf p} -
{\bf A}(t_1)\bigr]\Bigr\}\;,
\end{eqnarray}
where
\begin{equation}\label{w}
w = \tau\omega - {\bf r}\cdot {\bf  k} -
\int_{T-\tau/2}^{T+\tau/2}{d t_1\over\tau} {\bf r}\cdot {\bf A}(t_1)\;.
\end{equation}

It is illustrative to see how the static field results
\cite{highfield} can be recovered from Eqs.(\ref{A}-\ref{w}).
In
the limit $\Omega\to 0$ and for parabolic bands Eq.(\ref{A}) reduces to
\begin{figure}
\epsfxsize=7.5cm
\epsfbox{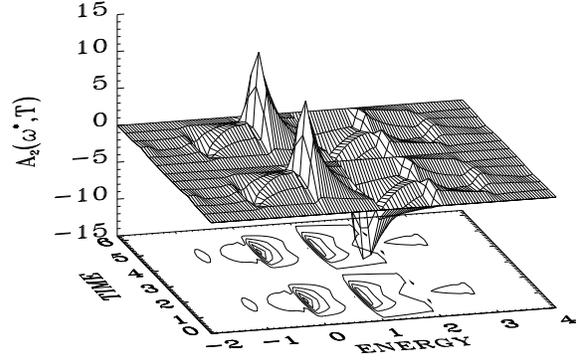}
\vspace{1cm}
\caption{The function ${\tilde A}_2$.  The parameters are defined
in the caption of Fig. 1.}
\label{f2}
\end{figure}
\begin{eqnarray}\label{Astatic}
A_{dc}({\bf k},\tau) & = & \exp[-i(\epsilon_k\tau + E^2\tau^3/24)]\nonumber\\
A_{dc}({\bf k},\omega) & = & (2\pi/\beta){\rm Ai}[(\epsilon_k-\omega)/\beta]\;,
\end{eqnarray}
with $\beta = (\hbar^2 e^2 E^2/8m)^{1/3}$, and Ai$(x)$ is the Airy
function. Evaluation of
the momentum sum leads then in three dimensions to
\begin{equation}\label{rhostatic}
\rho_{\rm dc}^{3D}(\omega) \propto
\Bigl[{ {\rm Ai}'}^2 \bigl ( -(\omega/\Theta)\bigr)
+ (\omega/\Theta){\rm Ai}^2\bigl ( -(\omega/\Theta)\bigr )\Bigr]\;,
\end{equation}
with $\Theta= 4^{1/3}\beta/\hbar$. Eq.(\ref{rhostatic}) is the
standard form for the Franz-Keldysh line-shape \cite{Haug-Koch}.

In the time-dependent case, one finds
\begin{equation}\label{Aharm}
{\tilde A}({\bf k},\omega,T) =
\int d\tau\cos\bigl[(\omega-\epsilon_k-\omega_F)\tau
+X(\tau) + 2 Y(\tau)\bigr]
\;,
\end{equation}
where we have introduced the notation
\begin{eqnarray}
X(\tau)& = & {\omega_F\over\Omega}\sin\Omega\tau\cos 2\Omega T
\\
Y(\tau) &= & {\omega_F\over\Omega}
{4\sin^2 (\Omega\tau/2) \sin^2 \Omega T \over\Omega\tau}\;,
\end{eqnarray}
and defined the field parameter
$\omega_F =e^2 F^2/(4 m \hbar\Omega^2)$\cite{interpret}.
An explicit analytic evaluation of the
integral in (\ref{Aharm}) is not possible, nor is
a straightforward numerical calculation feasible due to its
singular character.  The singularities, however, can
be isolated with analytic means, and we
proceed as follows.  The trigonometric
identity $\cos(x+y) = \cos x - 2\sin(x+y/2)\sin(y/2)$ allows
us to write (\ref{Aharm})
as a sum of two terms, ${\tilde A} = {\tilde A}_1
+{\tilde A}_2$, where
\begin{eqnarray}
{\tilde A}_1 ({\bf k}, \omega,T)  & =  & \int
d \tau \cos\bigl[(\omega-\epsilon_k-\omega_F)\tau
+X(\tau)\bigr]\;,\\
{\tilde A}_2 ({\bf k}, \omega,T)  & =  &
-2\int d \tau\sin\bigl[(\omega-\epsilon_k-\omega_F)\tau\nonumber\\
& & \quad\quad\quad + X(\tau) + Y(\tau)\bigr]\sin Y(\tau)\;.
\end{eqnarray}
\begin{figure}
\epsfxsize=7.5cm
\epsfbox{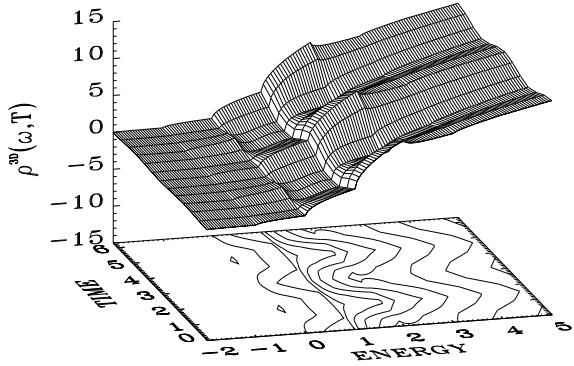}
\vspace{1cm}
\caption{Density of states, three dimensions.}
\label{f3}
\end{figure}
The integral defining ${\tilde A}_2$ is convergent,
and its numerical evaluation is straightforward. On the other hand,
analytic progress can be made with ${\tilde A}_1$,
which contains the aforementioned singularities.  Recalling the
identity
$
\int_{-\infty}^{\infty} d x \cos(ax-b\sin x) = 2\pi\sum_{n=-\infty}^{\infty}
J_n(b)\delta(n-a)\;,
$
where $J_n$ is the n-th order Bessel function,
we finally obtain
\begin{eqnarray}\label{A1}
{\tilde A}_1 ({\bf k}, \omega,T)  & = & 2\pi
\sum_{n=-\infty}^{\infty}
J_n[-{\omega_F\over\Omega}\cos(2\Omega T)]\nonumber\\
&\times&\delta(\omega - \epsilon_k -n\Omega-\omega_F)\;.
\end{eqnarray}

These two components ${\tilde A}_1$
and ${\tilde A}_2$ of the spectral
function have quite distinct properties.  The frequency
sum rule is exhausted by ${\tilde A}_1$ alone,
\cite{proof} {\it i.e.}
$\int d\omega/2\pi {\tilde A}_1(\omega,T) = 1$, while
$\int d\omega/2\pi {\tilde A}_2(\omega,T) = 0$.  Further,
the delta-function singularities of
${\tilde A}_1$ are reminiscent of {\it photonic side-bands}
\cite{photon},
however in the present case these features are shifted by the field- and
frequency-dependent parameter $\omega_F$.  In the
zero-field limit ${\tilde A}_1$ reduces to the field-free
result $a_0(k,\omega)=2\pi\delta(\omega-\epsilon_k)$, while
${\tilde A}_2$ vanishes.
Note also that the spectral function depends {\it quadratically} on
the electric field.  This is consistent with physical
intuition:  since there is no preferred direction,
it should not matter if the field direction
is reversed, ${\bf E} \to -{\bf E}$, and thus one
expects to find a spectral function that is even
in the applied field. The functions  ${\tilde A}_1$
and ${\tilde A}_2$ are displayed in Figs. 1 and 2, respectively.

We can use the above results to
calculate a time-dependent density-of-states function $\rho(\omega,T)$.
The contribution arising from ${\tilde A}_1$ is readily evaluated
by using the delta-function, while ${\tilde A}_2$ requires some
more work.  In three dimensions we find\cite{tricks}
\begin{eqnarray}
\rho_1^{3D}(\omega,T) & = & {(2m^3)^{1/2}\over
2\pi^2\hbar^{5/2}}\sum_{n=-\infty}
^{[{\omega-\omega_F\over\Omega}]}
J_n[-{\omega_F\over\Omega}\cos(\Omega T)]\nonumber\\
&\times&(\omega-\omega_F - n\Omega)^{1/2}
\end{eqnarray}
and
\begin{figure}
\epsfxsize=7.5cm
\epsfbox{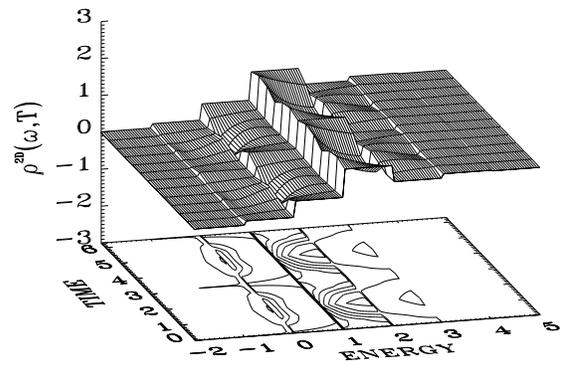}
\vspace{1cm}
\caption{Density of states, two dimensions.}
\label{f4}
\end{figure}
\begin{eqnarray}
\rho_2^{3D}(\omega,T) & = &{(\pi m^3)^{1/2}\over 4\pi^3\hbar^{5/2}}
\int_0^{\infty} {d\tau\over\tau^{3/2}}
\sin[Y(\tau)]\nonumber\\
&\times&
\sin[(\omega-\omega_F)\tau+X(\tau)+Y(\tau)+\pi/4]
\;.
\end{eqnarray}
\noindent
Here the summation over $n$ should be cut-off at the largest integer
not exceeding $[(\omega-\omega_F)/\Omega]$.
A numerical example is shown in Fig. 3.  The overall qualitative
picture is dominated by the free-electron like square-root components
generated by the photonic side-bands in ${\tilde A}_1$.  At different
points of time, however, the $n$:th side-band amplitude, which is determined
by the Bessel function $J_n$, may change its sign, which can be seen
in Fig. 3.  The more detailed fine structure
in $\rho^{3D}(\omega,T)$ is caused by the contribution
originating from ${\tilde A}_2$.

Similar, but even more dramatic effects can be found in
two dimensions where the result is
\begin{eqnarray}
\rho_1^{2D}(\omega,T) & = &{m\over2\pi\hbar^2}
\sum_{n=-\infty}
^{[{\omega-\omega_F\over\Omega}]}
J_n[-{\omega_F\over\Omega}\cos(\Omega T)]\nonumber\\
&\times&\theta(\omega -\omega_F - n\Omega)\\
\rho_2^{2D}(\omega,T) & = &{m\over 2\pi^2\hbar^2}
\int_{-\infty}^{\infty}{d\tau\over\tau}\sin[Y(\tau)]\nonumber\\
&\times&
\cos[(\omega-\omega_F)\tau+X(\tau)+Y(\tau)]\;.
\end{eqnarray}
The zero-field step-function reflecting the two dimensional density-of-states
is split into several substeps extending to energies well below the gap, and
the
above threshold values are modified by additional structures, whose amplitude
tends to zero for increasing energies.
Just as in the three-dimensional case, the relative amplitude of the
substeps changes as a function of time, and the fine structure in
the energy dependence is due to ${\tilde A}_2$.

It is also of interest to examine how the time-dependent
density of states depends
on the field strength at a fixed frequency.  The effects are most easily
analyzed by examining the time-average or the density of states,
$\rho_{\rm ave}(\omega)\equiv
(\Omega/2\pi)\int_0^{2\pi/\Omega}dT\rho(\omega,T)$,
and Fig. 5 shows the results of such a calculation.
\begin{figure}
\epsfxsize=7.5cm
\epsfbox{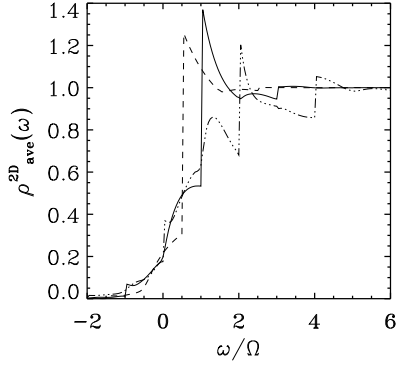}
\vspace{1cm}
\caption{Time-averaged
2-dimensional density of states (in units of zero-field
density of states) for $\omega_F /\Omega$= 0.5 (dashed line),
1 (solid line), and 2 (dash-dots).}
\label{f5}
\end{figure}
The following features of $\rho_{\rm ave}(\omega)$
are of importance: (i) The main
absorption edge moves to higher energies with increasing field
strength; (ii) There is a significant enhancement at the absorption
edge; and (iii) the range of energies where the edge is modified
increases with increasing field strengths.

Finally let us comment on the observability of the predicted effects.
There are several limiting factors: the spectral resolution in an
absorption measurement, the available frequencies, and the maximal
ac-field strengths that can be applied to the sample.  From the results
presented above we conclude that most favorable conditions
are obtained if $\omega_F/\Omega > 1$. It would appear that
this condition is most easily achieved at low frequencies. However,
a small $\Omega$ implies that all the fine-structure would be confined
to the immediate neighborhood of the absorption edge (in the example of
Fig. 5 the fine structure extends only a few $\Omega$'s above
the edge), and not resolvable
spectroscopically.  Increasing $\Omega$ leads to an increased $\bf E$ in
order to maintain sufficiently large $\omega_F/\Omega$.  By using parameters
attainable with free-electron lasers\cite{FEL}, $\Omega\simeq$ 1 THz,
$ {\bf E}\simeq$
1 MV/m, and the GaAs effective mass, we estimate that $\omega_F/\Omega\simeq
1$.
This implies that the fine-structure extends a few meV around the
absorption edge, and should be observable.

In summary, we have calculated the density of states of free carriers in
time-dependent fields, and find that a harmonically varying external field
leads to a field and frequency dependent shift of the
main absorption edge, and gives rise to
fine structure (``dynamical Franz-Keldysh effect") of
observable magnitude.

We acknowledge useful comments from Prof. S. Koch and Prof. Jai Singh.


\begin{references}
\bibitem{dimensions}
The analysis following Eq.(\ref{Dyseq}) is independent of the number of spatial
dimensions, however
one should be aware of the tacit assumption that the electric field
lie in a direction in which there is translational invariance.
\bibitem{interpret}
The parameter $\omega_F$ has an appealing physical interpretation:
$\hbar\omega_F$ is the time-averaged kinetic energy of a free particle
with charge $e$ and mass $m$ in a harmonic electric field of
frequency $\Omega$.
\bibitem{proof}
The proof is based on the properties of Bessel functions:
(i) $J_{-n}(z)=(-1)^nJ_n(z)$, and (ii) $1 = J_0(z) + 2J_2(z)
+ 2J_4(z) + \dots$.
\bibitem{tricks}
The momentum summations can be performed using the
formal relations $\int_0^{\infty}dx \sqrt{x}\exp[\pm i a x]
=$ $(-1\pm i)\sqrt{2\pi}/(4a^{3/2})$ (three dimensions), and
$\int_0^{\infty} dx \exp[\pm i a x] = \mp 1/(i a)$ (two dimensions),
which can be used in the distribution sense within the integrals
needed for the density of states.
\end{references}
\end{document}